# High thermoelectric power factor in Ni-Fe alloy for active cooling applications

Shuai Li,[a] Sree Sourav Das,[b] Haobo Wang,[a] Sujit Bati,[c] Prasanna V. Balachandran,[a] Junichiro Shiomi,[d] Jerrold A. Floro [a] and Mona Zebarjadi [a,b]

a. Department of Materials Science and Engineering, University of Virginia, Charlottesville, VA, 22904, United States
b. Electrical and Computer Engineering Department, University of Virginia, Charlottesville, VA, 22904, United States
c. Department of Physics, University of Virginia, Charlottesville, VA, 22904, United States
d. Department Of Mechanical Engineering, University of Tokyo, Tokyo, Japan

Metallic thermoelectric materials are promising candidates for active cooling applications, where high thermal conductivity and a high thermoelectric power factor are essential to maximize effective thermal conductivity. While metals inherently possess high thermal and electrical conductivities, they typically exhibit low Seebeck coefficients. In this work, we create a database and apply machine learning techniques to identify metallic binary alloys with large Seebeck coefficients. Specifically, we identify Ni-Fe as a promising candidate for active cooling. We then fabricate Ni-Fe ingots and demonstrate thermoelectric power factor values as high as 120 μW/cm·K² at 200 K for these stable alloys, which are composed of cost-effective and abundant elements. Furthermore, we show that the effective thermal conductivity of these alloys, under small temperature differences, can exceed that of pure copper at temperatures above 250 K.

## Introduction

With the increasing density of transistors and operating frequencies in integrated circuits (ICs), driven by rapid advancements in semiconductor technologies, efficient heat dissipation has become an increasingly critical challenge. Inadequate heat management impairs the performance of these densely packed circuits and jeopardizes their reliability. Conventional cooling techniques, such as passive heat sinks and fluid-based cooling systems, often struggle to meet the efficiency, size, and design requirements of ICs. In response, novel approaches—such as active cooling based on the thermoelectric (TE) effect—have emerged as promising solutions for thermal management.

TE materials, known for their ability to convert thermal gradients into electrical energy and vice versa, offer versatile applications through both the Seebeck and Peltier effects. These materials have long been studied for power generation and refrigeration, with the performance of TE devices governed by the dimensionless figure of merit, $zT = \frac{\sigma \alpha^2}{\kappa}T$, where σ is the electrical conductivity, α is the Seebeck coefficient, T is the temperature, and $\kappa$ is thermal conductivity. We note that $\kappa$ is the passive thermal conductivity in the absence of electric current. Improvement of $zT$ requires strategies to increase the TE power factor ($PF = \sigma \alpha^2$) and/or to decrease the $\kappa$.[1–5] However, recent advances have expanded the role of TE materials to include active cooling modes[6–9], where the Peltier current can actively enhance the passive heat transfer. Under optimum current conditions and when a TE module with a length L is placed between a hot object characterized by $T_H$ and a cold heat sink at $T_C$, Peltier cooling can be expressed as $J_{peltier} = \frac{\sigma \alpha^2 T_H^2}{2L}$. One can therefore combine passive and active heat flux where this unique mode of operation, is characterized by the concept of effective thermal conductivity $\kappa_{eff} = \left(\kappa + \frac{\sigma \alpha^2 T_H^2}{2\Delta T}\right)$, combines the passive ($\kappa$) and active (Peltier) components to maximize the heat flux, opening new opportunities for TE application in thermal management.

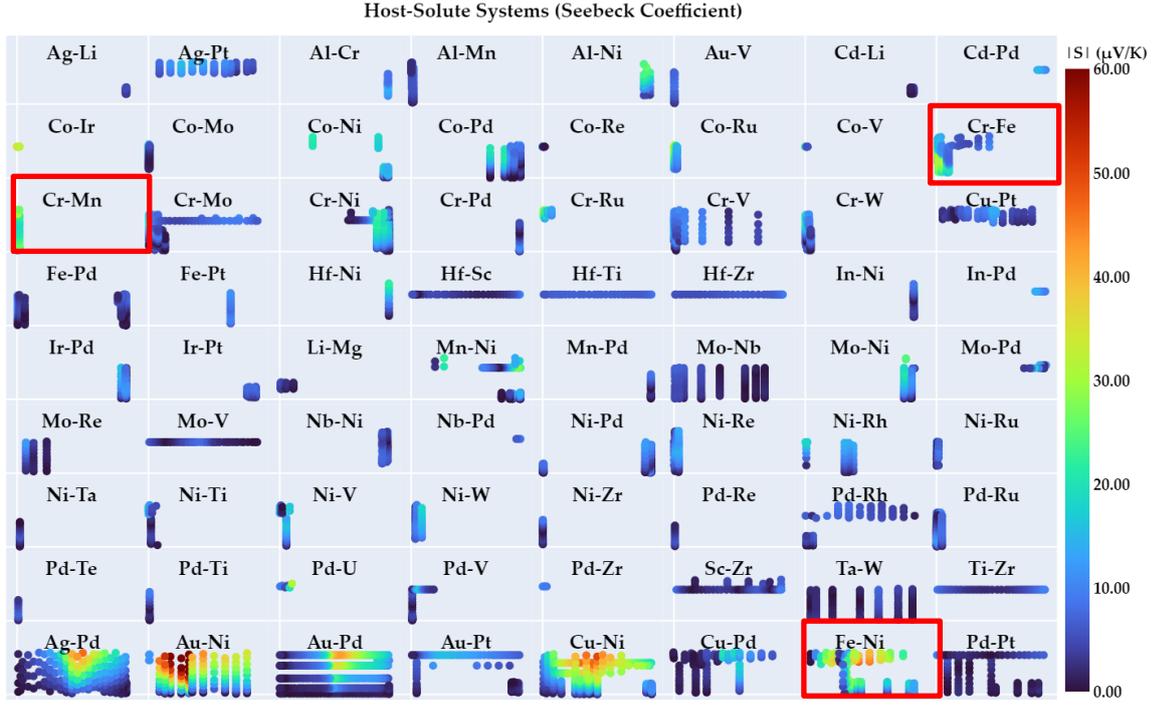

Figure 1. Visual representation of part of our database. Seebeck values of less than a few microvolts per Kelvin are eliminated and the rest are represented here. Each subplot represents the host-solvent binary alloy labelled in the plot. For example, Cu-Ni means Cu is the host, Ni is the solvent. Each subplot has an x-axis scale of 0 to 1 representing the molar fraction of the solvent. The y-axis is the measurement temperature from 0-400 K for all subplots. The colour represents the absolute value of the Seebeck coefficient. Promising yet less-explored candidates are highlighted with red rectangular boxes.

Traditional TE materials with low $\kappa$ are not suitable for active cooling. Instead, metallic TE materials are promising due to their inherently high electrical and thermal conductivities, originating from the high concentration of free electrons, n. The Seebeck coefficient of metals is in general lower than that of semiconductors and is explained by the Mott Formula.

$$S = -\frac{\pi^2 k_B^2 T}{3q\sigma(\mu)} \frac{d\sigma(E)}{dE} \Big|_{E=\mu} \qquad (1)$$

Where $k_B$ is the Boltzmann constant, T is the temperature, q is the charge of the electron, E is energy, $\mu$ is the chemical potential, $\sigma$ is differential conductivity or transport function where $\sigma(E) = \frac{q^2}{3} DOS(E) v^2(E) \tau(E)$, $DOS$ is the density of states, $\tau$ represents the relaxation time, and $v$ is the group velocity. A high *DOS* at the Fermi level leads to an increased carrier density, which in turn enhances both electrical and thermal conductivity. However, according to the Mott formula (Eq. 1), a large *DOS* is associated with a reduced Seebeck coefficient. Near room temperature, a large list of pure metals, including Ag, Al, Au, Cd, Cs, Cu, Dy, In, Ir, Mg, Nb, Pb, Rh, Sn, Sr, and Ru, shows absolute Seebeck below 5 $\mu V/K$.[10] A few elemental metals e.g. Co[11], Fe[12], and Ni [13] have larger Seebeck coefficient values (|S| ~ 20$\mu V/K$).[10] In the case of Ni, the large Seebeck coefficient can be attributed to the sharp slope of the *DOS* at the Fermi energy due to partially filled d-orbital[14], which leads to a high Seebeck (Eq. 1). Due to the inherent magnetization of these elements (Ni, Co & Fe), part of their Seebeck coefficient has been attributed to the magnon-drag effect wherein the magnon heat flux drags along the electronic charge carriers. Watzman et al. [15] discussed two magnon-drag contributions to the Seebeck coefficient: the hydrodynamic contribution and the spin-motive force contribution. They demonstrated that magnon-drag is the dominant component of the Seebeck coefficient of iron and cobalt.

Binary alloys of transition metals are shown to be good candidates to further increase the TE power factor of metallic systems, examples include Cu-Ni[9,16], Au-Ni[14], Fe-Ni[17,18], Cr-Mn[19], Pd-Ag[20], and Cr-Fe[21]. A summary of previously reported Seebeck coefficients of binary metallic systems up to 400 K is shown in Figure 1, the majority of which is derived from the Landolt Bornstein database.[22] In several cases, the Seebeck coefficient and the power factor of the alloy are larger than both parent elements which can be attributed to either changes in the density of states or modifications of the scattering rates. Here we highlight three of the most studied solid-solution alloys, Pd-Ag, Cu-Ni, and Au-Ni. In all three cases, the alloy demonstrates a higher Seebeck coefficient compared to both the host and solute. The Pd-Ag alloy exhibits a peak Seebeck coefficient of approximately 40 µV/K at 300 K reaching ~ 80 µV/K at 1300 K with 55% Pd.[20] These values are larger than pure Pd which has a negative Seebeck coefficient in the 300K to 1300K range, and pure Ag with positive values below 10µV/K. However, the limited availability of Pd and the cost of the elements restricts its widespread use. Constantan (Cu-Ni) alloy is composed of abundant and low-cost elements and is easy to synthesize.[23] Constantan is reported to have a TE power factor of 40 $\mu W\ cm^{-1}\ K^{-2}$ at 300 K and 102 $\mu W\ cm^{-1}\ K^{-2}$ at 873 K.[16] Constantan is also studied in the context of active cooling and using additive manufacturing for industrial applications.[9] Au-Ni in contrast is expensive and metastable. However, due to its very large TE power factor worthy of investigation. Garmroundi et al.[14] reported a Seebeck coefficient of $94 \mu V/K$ for a quenched, metastable single face-centered cubic (FCC) Ni-Au alloy at 1000 K, resulting in an ultra-high peak power factor of $340 \mu W\ cm^{-1}\ K^{-2}$ in $Ni_{0.1}Au_{0.9}$ sample at 560K. This large power factor is hypothesized to arise from the selective scattering of s-electrons into localized d-states, which induces strong energy-dependent scattering rates $\tau(E)$ and enhances the slope of $\sigma(E)$ near the Fermi-level, thereby increasing the Seebeck coefficient (see Eq. 1).

Ni-Fe alloys also have been studied due to their significance in geology, meteoritics, and material science.[24–32] There are old and scattered studies reporting the Seebeck coefficient data of Ni-Fe alloys as a function of compositions and temperatures [17,18,33], showing the peak Seebeck of Ni-Fe alloys can reach -50 µV/K with 53.8 at% at 300 K [18] and -46 µV/K with 40 at%[17] of Ni concentration at 200 K. Figure 2(c-d) highlights that while the Fe-Ni system has been extensively studied between 300 and 1200 K, only a limited number of compositions have been investigated below 300 K. Similarly, Cr-Mn and Cr-Fe alloys were also examined but within a specific compositional range, as depicted in Figures S2(a) and S2(b), respectively. This research gap drives our focus on these three binary systems, aiming to enhance their performance through further composition optimization from low temperatures to room temperature ranges and explore the potential alloys for active cooling applications.

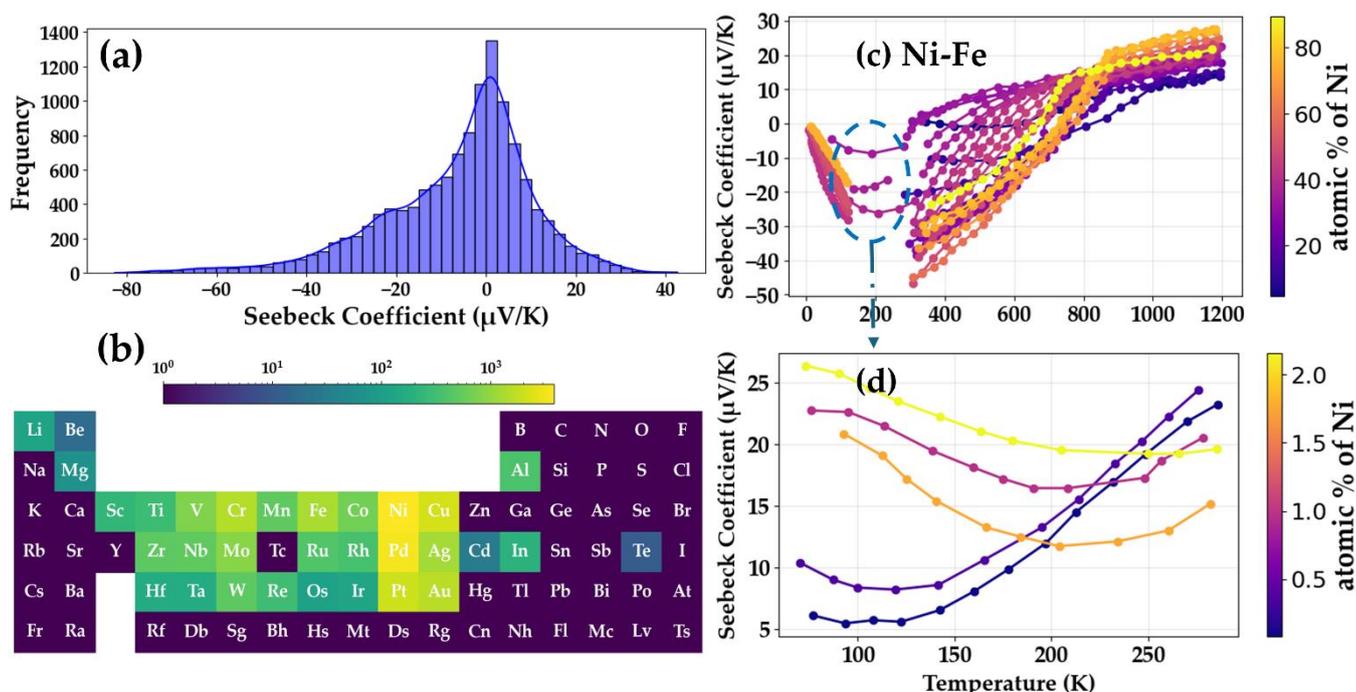

Figure 2. (a) Distribution of Seebeck coefficient in the database, (b) Frequency of the elements in the dataset, for example, Ni and Pd are the most frquently used metals, and (c) Temperature dependent Seebeck coefficients of Fe-Ni system upto 1200K with respect to different atomic % of Ni, (d) highlighting low temperature region from 50 K to 300 K.

In the past decades, materials optimization has primarily relied on traditional trial-and-error methods, guided by the physical and chemical intuition of scientists. However, these approaches are often time-consuming, inefficient, and costly.[34] To accelerate the optimization process, we employ a machine learning (ML) approach. By utilizing a data-driven model, our method efficiently explores a broad range of metallic compounds, focusing on different atomic concentrations, temperatures, and predicted Seebeck coefficients. In what follows we present our material database that we built based on binary solid-solution metallic alloys. Using this database, we selected and optimized these three alloys and finally, validated the results experimentally.

**Selection and Optimization using Machine Learning:**
**Dataset**
Our initial dataset was obtained from the Landolt-Börnstein database[22] and a recent publication on Au-Ni alloys[14]. It comprises experimental Seebeck coefficient values for various binary metallic systems, recorded in wide temperature and atomic concentration ranges. The data was extracted and digitized manually using the GRABIT MATLAB tool, resulting in a total of 12,332 data points with 3,103 unique solid solutions. The Seebeck coefficient values span from 40 to -85 µV/K, covering a temperature range of 0 to 1500 K as depicted in Figure 2 and Figure S1(a), respectively. [9]

Figure 2(a) demonstrates the distribution of the Seebeck coefficient in this dataset revealing that most metals have small Seebeck values (a few µV/K) which is one of the main reasons for the TE community to focus on semiconductors instead. Figure S1(b) provides an overview of the metals included in the dataset, showing that Ni is the most frequently used material. It is followed by Pd, Cr, Pt, and Cu. This information is presented in an alternative format in Figure 2b.

**Feature Selection**
To build generalizable data-driven models, it is important to include features that not only capture the trend of Seebeck coefficients across different metallic alloys but also uniquely represent them. In this regard, a Composition-Based Feature Vector (CBFV)[35] technique was used to derive features from the chemical formula, utilizing the Materials Agnostic Platform for Informatics and Exploration (Magpie). [36] Furthermore, temperature, and crystallinity information (Single-crystal /Polycrystalline) by level encoding (1/0) are also added to the feature list which turns in a total of 156 input features and Seebeck coefficient as the target value. A detailed table summarizing the input features is provided in Table S1. These features are commonly used in the field of materials informatics of TE.[37–41] The correlation analysis, as shown in Figure S3, indicates that many features exhibit strong statistical correlation[1]. In most cases, it is advisable to remove one of two highly correlated features since they convey redundant information. Such correlations can hinder model convergence, degrade predictive performance, and affect interpretability.

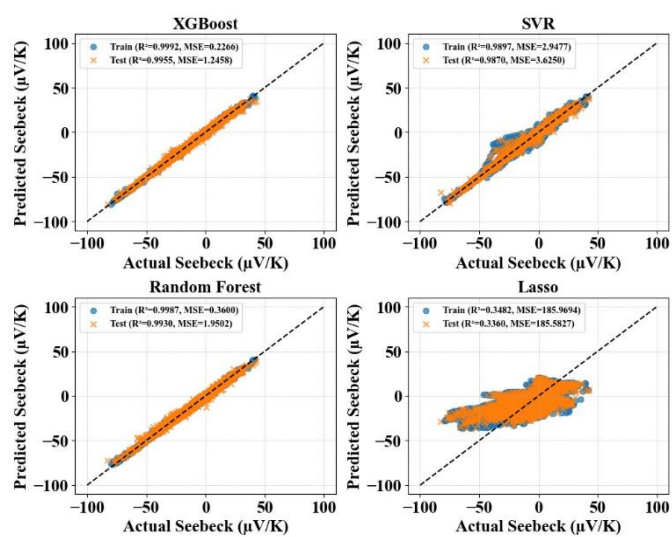

Figure 3. Predicted vs Actual Seebeck coefficients from XGB, SVR, RF and Lasso. Tree and Kernel-based models outperformed the linear model

The dimensionality of the input features was reduced by applying a correlation coefficient threshold of 0.5.[42,43] This means that only those features with an absolute correlation coefficient less than 0.5 with other features were kept to ML model building. This process reduced the number of input features to 19 which were used to predict the Seebeck values.

**ML Model:**
We employed three distinct types of ML models: a linear model based on Least Absolute Shrinkage and Selection Operator (LASSO)[44] regression, tree-based models (Extreme Gradient Boosting (XGBoost)[45] and Random Forest (RF)[46], and a kernel-based model (Support Vector Regression (SVR)[47]. Based on the accuracy-interpretability trade-off[48], linear models offer higher interpretability, while kernel-based models provide greater accuracy at the cost of interpretability. Tree-based models fall between these two extremes, balancing accuracy and interpretability.

**Training and Testing of the Models:**

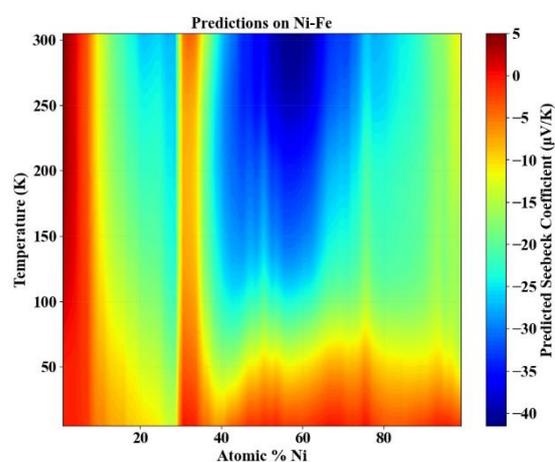

Figure 4 Heat map of Ni-Fe alloy averaging over all predictions from all optimized models.

To avoid any perceived bias during training, we employed a data-driven approach for splitting the dataset. We performed K-means [49] clustering analysis (using Euclidean distance as the similarity metric) on the dataset, and the Silhouette score, as shown in Figure S4, suggests that the dataset contains two distinct clusters. Each cluster represents a different group of data points that share common characteristics. Cluster 1 consists of 8,743 data points, while Cluster 2 contains 3,589 data points. To ensure that the models learn from both types of data distributions, we randomly selected 70% of the data from each cluster to form the training set (8,632 data points), with the remaining 30% used as the testing set (3,700 data points). This approach ensures a more balanced representation of samples from both clusters in the training and testing sets.

**Hyperparameter Tuning:**
Hyperparameter tuning helps improve model performance and prevents overfitting. To optimize the hyperparameters, we used BayesSearchCV from the scikit-optimize library[50] in Python.

BayesSearchCV employs a Gaussian Process Regression as a surrogate model for hyperparameter optimization. An acquisition function is used to determine which hyperparameter combinations to evaluate next, with Expected Improvement (EI) as the default acquisition function. The EI function estimates the expected improvement over the current best result.[51] During optimization, 10-fold cross-validation from scikit-learn library [52] was applied, which partitioned the training set into ten subsets. Each model is trained in nine subsets and validated on the remaining subset. The testing set remained unseen by the models during cross-validation. The lower and upper boundaries for the hyperparameters are provided in Table S2. The number of iterations for the optimization process was set to 50. The optimized hyperparameters for each model in each case of splitting are shown in Table S2. The performance of these models was evaluated by comparing two key metrics, namely the coefficient of determination ($R^2$) and mean squared error (MSE).

**ML Prediction:**
Figure 3 presents a comparative analysis of the predicted vs. actual Seebeck coefficients across different models. The low $R^2$ and high MSE values in both the training and test sets for LASSO LASSO indicate that the linear model failed to capture the complex relationships between the input features and Seebeck coefficients. In contrast, XGB, RF, and SVR demonstrated a strong predictive performance, achieving an $R^2$ of 0.99 on both training and independent test sets, highlighting their ability to model the non-linear dependencies. After successfully evaluating the prediction performance of the models on the test data, we applied the optimized models (XGB, RF, and SVR) to predict the Seebeck coefficients of Ni-Fe, Cr-Mn, and Cr-Fe alloys across the entire compositional range from 50 to 305 K. The final Seebeck coefficient values were determined by averaging the predictions from all three models, with detailed predictions provided in Table S3. Based on these averaged results, the Ni-Fe system exhibited the highest Seebeck coefficient of 42.3 µV/K at 59 at% Ni and 305 K. Similarly, the Cr-Mn system reached a peak Seebeck coefficient of 34.6 µV/K at 10 at% Mn and 305 K, while the Cr-Fe system achieved 36.4 µV/K at 2 at% Fe and 155 K. Heat maps illustrating the Seebeck coefficients of the Cr-Mn and Cr-Fe systems are shown in Figure S5. Among these three systems, Ni-Fe presents the highest Seebeck coefficient. In addition, experimental validation of Cr-based systems presents challenges due to chromium's high reactivity with steel milling jars, which can alter the sample composition and degrade the performance.[53,51] Therefore, in this study, we focus on the Ni-Fe binary alloy system, given its high Seebeck coefficient near room temperature as predicted by the ML models. Additionally, this system is notable for its low cost, scalability for industrial applications, ease of synthesis, and exceptional mechanical durability.[54,55] These characteristics make Ni-Fe alloys highly suitable for practical TE device applications. According to the heat map of the predicted Seebeck coefficients (Figure 4), while the peak composition is $Ni_{56}Fe_{44}$, the alloy retains a Seebeck coefficient above -40 µV/K within the 48–62 atomic % Ni range around from 250 K to room temperature. This indicates a compositional window where the material could exhibit high TE performance. By identifying the peak Seebeck coefficient and compositional range, the ML model provides a focused direction for experimental exploration, minimizing the need for testing across all possible compositions.

In what follows we present our experimental approach to the power factor and effective thermal conductivity characterization of the Ni-Fe alloys in the 50-400K temperature range inspired by the ML predictions. Further, we present the microstructure of the as-arc-melted sample and the homogeneity of the solid-solution alloy at microscales for active cooling applications.

**Experimental Methods**

Iron powder with 99.5% purity and nickel powder with 99.996% purity were weighed to 5 grams per sample and mixed in an argon-filled glovebox. The powder mixtures were then hot-pressed into solid bulk samples at 800°C under 56 MPa pressure for 300 seconds using an OTF-1700X-RHP4 hot-press setup from MTI Corporation. The

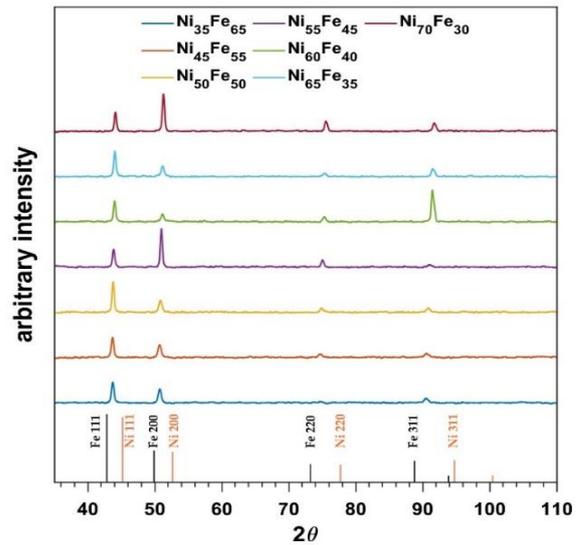

Figure 5. XRD results of arc-melted Ni-Fe samples. The black and orange straight lines on the bottom are FCC iron and nickel reference peaks.

solid bulk samples were later arc-melted to form a Ni-Fe solid solution. Each sample was melted and flipped twice for homogeneity and then melted without flipping allowing bubbles and voids to diffuse to the top of the sample which was then cut out. The central-bottom part of the arc-melted samples was then sectioned into approximately 2mm×2mm×10mm bar shape. Transport properties were measured using the Thermal Transport Option of Quantum Design PPMS Versalab. A heater was attached to one side of the sample to create a 3% rise in temperature. The other side was connected to a heat sink. The resulting voltage difference and temperature difference under steady state were measured along the length of the sample to extract the Seebeck coefficient and the thermal conductivity. The heater and the heat sink contact (copper coated with gold) were then used to send current along the sample. The voltage was measured using side probes enabling 4-probe electrical

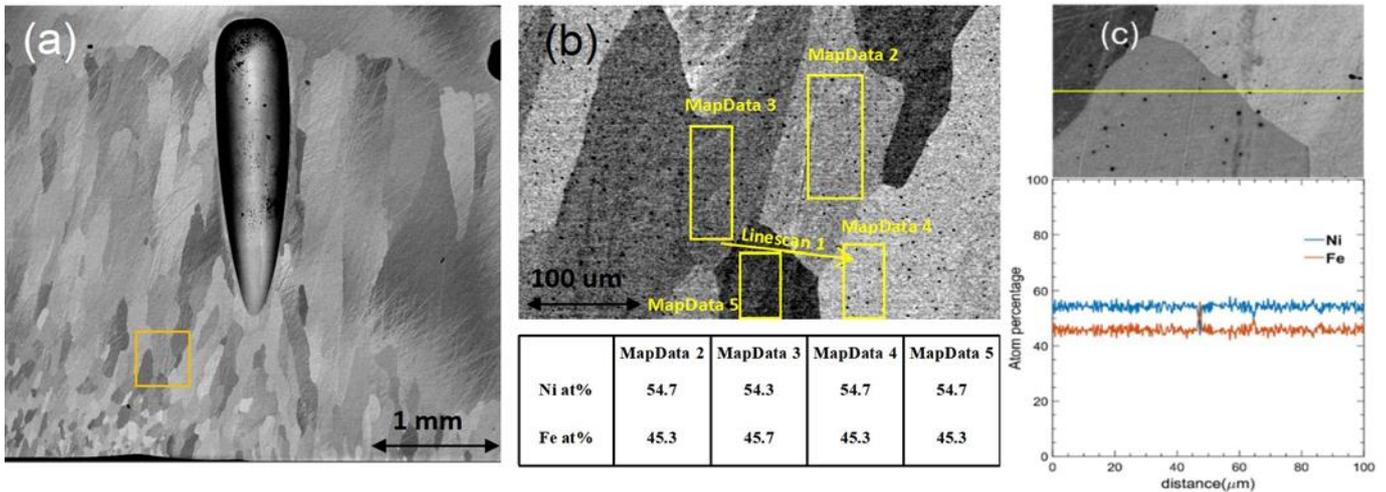

Figure 6. SEM backscattered electron image on the cross-section of the arc-melted $Ni_{55}Fe_{45}$ sample(a), the yellow square marked area is the chosen EDS mapping area. This area is enlarged in (b) where the EDS mapping locations on different grains are reported and summarized in the table. (c) EDS line scan across the grains over the line scan 1 shown in (b)

conductivity measurements. The XRD characterization is performed using an Empyrean X-ray diffractometer from Malvern-Panalytical on the sectioned as-arc-melted ingots. SEM/EDS is performed on an FEI Quanta 650 Scanning Electron Microscope (SEM).

## Results and Discussion

### Material Characterization

According to the phase diagram, Ni-Fe alloys form an FCC Ni-Fe solid solution within the probed composition range at elevated temperatures. Due to slow diffusion [27–30,56], decomposition of g to a + FeNi$_3$ is unlikely under our experimental conditions, resulting in g phase solid-solution.

The X-ray diffraction (XRD) data of all Ni-Fe alloy samples are shown in Figure 5. The bulk XRD measurements are performed on the vertical cross-section of the as-arc-melted ingots. Going from bottom to top, Ni concentration in the samples increases.

All samples exhibit a consistent series of diffraction peaks, corresponding to the (111), (200), (220), and (311) crystallographic planes of the FCC structure. The variation of intensities of peaks among samples can be attributed to non-random distributions of

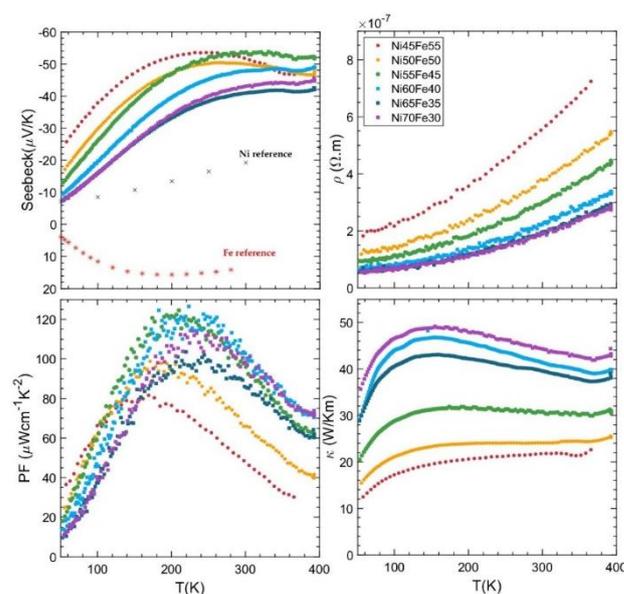

Figure 7 (a) Seebeck coefficient, (b) resistivity, (c) power factor, (d) thermal conductivity of Ni-Fe samples. The black and red crosses in (a) are reference Seebeck of pure Ni and Fe grain orientations at the section surface, which is confirmed by the SEM results. The orange and black lines at the bottom indicate the reference peak positions for pure Ni and Fe in the FCC structure. The peaks of the alloys are positioned between the reference peaks for pure Ni and Fe, indicative of solution formation. As the Ni concentration increases, the peaks shift to higher 2θ values, indicating smaller lattice parameters.[57] The XRD results confirmed that the prepared Ni-Fe samples are single-phase polycrystalline.

Figure 6 presents the backscattered electron image of the vertical cross-section of the as-arc-melted Ni$_{55}$Fe$_{45}$ sample. The different greyscale regions represent the varying crystal orientations of the grains, demonstrating the polycrystalline nature of the sample. At the top of the sample, a prominent needle-shaped bubble is observed, likely formed from the degassing of the powders during the melting process. Several smaller bubbles can also be seen in the upper section of the cross-section. To avoid these bubbles, subsequent transport measurements were performed on samples cut from the center of the lower portion where grains are more uniform.

The SEM image clearly illustrates the distribution of grain size and shape across the sample. The bottom of the sample, which was in contact with the water-cooled copper plate of the arc-melter, experienced a higher cooling rate, resulting in smaller grains (~ 100 μm). In contrast, the upper section contains large, elongated grains measuring up to several millimeters in length. Further EDS mapping of the highlighted area was conducted to characterize the composition of different grains. The black dots visible in the enlarged image are colloidal silica residues from the sample polishing process. The atomic composition data in Figure 6b summary table confirms the homogeneous composition across the grains, consistent with the stoichiometric ratio of the starting powders. As shown in Figure 6c the line scan reveals that the composition remains uniform both within and across grains. Additional SEM/EDS characterizations (Supplementary) performed on different samples and in different areas and orientations support that the samples are homogeneous and consistent with the measured composition.

**Thermoelectric Transport Properties**

Figure 7 summarizes the TE measurements performed on the arc-melted Ni-Fe alloy samples.

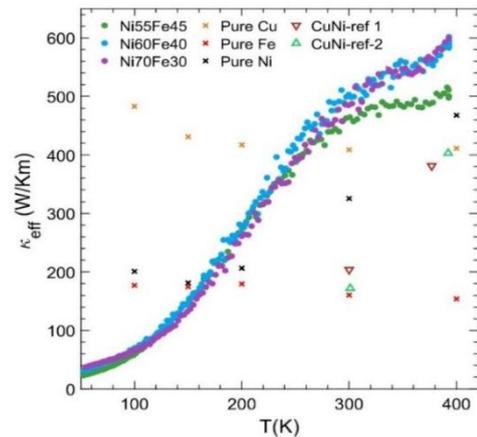

Figure 8. Effective thermal conductivity of Ni-Fe alloys compared with pure Cu, pure Fe, pure Ni, and reference Cu-Ni alloys: twin-boundary enhanced Cu-Ni[42] and ball mill-hot pressed Cu-Ni[43].

Alloys with a composition range of 45 to 70 atomic % Ni have an absolute value of the Seebeck coefficient which is up to 2.5 times greater than that of pure Ni or Fe. The peak Seebeck coefficient varies with composition, with the highest values observed in $Ni_{55}Fe_{45}$ and $Ni_{45}Fe_{55}$, both reaching -52 µV/K. This is consistent with the ML prediction presented earlier where 45-55% Ni was identified as the composition with the highest Seebeck value. The Seebeck coefficient's dependence on composition changes with temperature: at lower temperatures (<200K), the absolute value of the Seebeck coefficient decreases with increasing Ni content. However, this trend does not hold at intermediate temperatures (200K to 400K).

A previous work[18] observed a similar concentration dependence of the Seebeck coefficient. They attributed the trend at the higher temperatures to the concentration fluctuation within their samples, which is not supported by the SEM/EDS results in this paper.

We have shown the details of DOS calculations in supplementary materials wherein we have shown the slope of the DOS at the Fermi level does not correctly predict the sign of the Seebeck coefficient. Hence energy-dependent scattering rates are needed along with magnon-drag contributions to fully understand the Seebeck values of this alloy system. The resistivity increases with higher Fe content. The resistivities of the Ni-Fe alloys, ranging from $5.60 \times 10^{-8}$ Ω·m to $7 \times 10^{-7}$ Ω·m, highlighting the highly metallic nature of these alloys. Combining high Seebeck coefficients for these alloys with their low resistivity, the peak power factor reaches 120 µW/cm·K² for both $Ni_{60}Fe_{40}$ and $Ni_{55}Fe_{45}$. This is 20% higher than the peak values reported at 750K in previous studies on Cu-Ni alloys. In temperature ranges slightly below room temperature (i.e. 200K to 300K), there are not many candidates with extremely large TE power factors (i.e. above 100 µW/cm·K²). Commonly used TE materials in this temperature range include bismuth-tellurium-antimony-selenium-based materials generally have power factor values well below 100 µW/cm·K², with a recent work highlighting a record high value of 63 µW/cm·K² in this class of materials.[58] Au-Ni is reported to have a power factor slightly below 300 µW/cm·K² however, its cost and instability are not favorable. Single crystal $YbAl_3$ has a power factor slightly below 200 µW/cm·K² at room temperature. Other examples include low dimensional materials such as nm thin $FeSe$[59] and 1D $Ta_4SiTe_4$ samples[60]

At 200K, the power factors of $Ni_{60}Fe_{40}$ and $Ni_{55}Fe_{45}$ are larger than those of the hot-pressed $YbAl_3$ sample[61] and are much larger than that of the Cu-Ni alloy[16]. However, the power factor values decrease rapidly with increasing Fe due to the increase in resistivity and with Ni concentration due to the reduction in the Seebeck coefficient. Since the TE power factor is our primary focus, the Ni-Fe composition range is restricted to 45% to 70% atomic Ni. In this range, the thermal conductivity generally increases with Ni content.

The effective thermal conductivity ($κ_{eff}$) of the $Ni_{60}Fe_{40}$ sample exhibits the best balance between power factor and thermal conductivity. As shown in Figure , the $κ_{eff}$ of the Ni-Fe samples under a 1K temperature gradient is 2 to 3 times higher than the $\kappa_{eff}$ of pure Fe or Ni in the above 200K range. Above room temperatures, $\kappa_{eff}$ of the $Ni_{60}Fe_{40}$ alloy is still higher than that of pure copper and previous studies of high power factor Cu-Ni alloys[62,63], reaching 600 W/m.K for both $Ni_{60}Fe_{40}$ and $Ni_{70}Fe_{30}$ alloys.

As indicated by SEM, in the arc-melted samples, grains are significantly larger than the typical electron and phonon mean free paths in metals[64–66], eliminating the possibility of grain size influencing TE properties, especially the Seebeck coefficient. However, given the limited studies on Ni-Fe alloys as TE materials, further investigations with improved parameter control are essential to elucidate the role of microstructure in the thermoelectric performance of these alloys.

**Conclusion**

In summary, we built a database of binary metallic alloys and identified Ni-Fe as a potential candidate for active cooling applications. We used ML algorithms to identify the best molar fraction corresponding to the largest Seebeck values in the 45%-55% Ni range. We then proceeded with experimental validation of this prediction. The highest Seebeck values were observed in $Ni_{55}Fe_{45}$ and $Ni_{45}Fe_{55}$ samples consistent with ML prediction. The power factor and effective thermal conductivity of arc-melted Ni-Fe alloys with 45 to 70 atomic percent nickel were investigated over the 50K to 400K temperature range. Notably, the $Ni_{55}Fe_{45}$ and $Ni_{60}Fe_{40}$ alloys demonstrated a large peak power factor of 120 µW/cm·$K^2$ at 200K. This metallic binary alloy is stable and is composed of cost-effective and abundant elements. The power factor value reported is one of the largest values reported in this temperature range. The effective thermal conductivity, $\kappa_{eff}$, at a 1K temperature difference was also calculated using the measured values of passive thermal conductivity and TE power factor. The largest $\kappa_{eff}$ values exceeding 600 W/K·m at 400K were observed for $Ni_{60}Fe_{40}$ and $Ni_{70}Fe_{30}$ alloys outperforming pure copper, Ni, Fe, and state-of-the-art Cu-Ni alloys under the same conditions. The microstructure of the arc-melted Ni-Fe ingots was characterized using SEM and EDS, providing insights into grain size and elemental distribution. The abnormal composition dependence of the absolute Seebeck coefficient at intermediate temperatures (200K-400K) was also noted. A hypothesis suggesting that local concentration fluctuations account for this anomaly was tested using EDS analysis, which invalidated this explanation. Further research is needed to assess the effects of grain size, magnetic domains, and defects on the thermoelectric performance of Ni-Fe alloys. This study reveals the overlooked potential of Ni-Fe alloys for high-power factor applications, highlights the promise of transition metal alloys in the search for high-power factor metallic materials, and encourages further research into metallic thermoelectric materials for active cooling.

**Author contributions**

Shuai Li: Methodology and formal analysis (sample preparation, XRD, transport measurements), writing-original draft; Sree Sourav Das: Methodology, data curation, and formal analysis (database, ML), writing-original draft; Haobo Wang: Methodology (sample preparation, SEM), Sujit Bati: Methodology (DFT calculation), Prasanna Balachandran: Formal analysis (ML algorithms), writing-review & editing; Junichiro Shiomi Formal analysis (ML algorithms), writing-review & editing; Jerry Floro: Formal analysis, supervision, writing-review & editing, Mona Zebarjadi, conceptualization, supervision, writing-original draft, Funding acquisition

**Conflicts of interest**

There are no conflicts to declare.

**Data availability**

Data for this article, raw SEM and EDS scans are available at https://doi.org/10.18130/V3/AOSTG4. The dataset containing Seebeck coefficients for binary metallic systems will be available soon **in** figshare data repository.

**Acknowledgments**

This work is supported by NSF grant number 2421213. MZ and SSD acknowledge discussions with Masato Onishi and Ryo Yoshida on the details of the ML algorithm. SSD and SB acknowledge the Rivanna cluster of UVA used for the computational part.